\documentclass[conference]{IEEEtran}

\usepackage{amssymb}
\usepackage{amsmath}
\usepackage{amsthm}
\usepackage{mathtools}
\usepackage{graphicx}
\usepackage{caption}
\usepackage{subcaption}
\usepackage{url}
\usepackage{cite}
\usepackage[T1]{fontenc}
\usepackage{array}
\usepackage{color}
\usepackage{soul}
\usepackage{multirow}
\usepackage[utf8]{inputenc}
\usepackage{graphicx}
\usepackage{lineno}

\IEEEoverridecommandlockouts

\DeclareMathOperator*{\argmax}{\arg\!\max}

\graphicspath{{./figures/}}

\begin{document}

\title{Phase Identification in Distribution Networks with Micro-Synchrophasors}

\author{\IEEEauthorblockN{Miles H.F. Wen\IEEEauthorrefmark{1}, Reza Arghandeh\IEEEauthorrefmark{2}, Alexandra von Meier\IEEEauthorrefmark{2}, Kameshwar Poolla\IEEEauthorrefmark{2}, Victor O.K. Li\IEEEauthorrefmark{1}}
\IEEEauthorblockA{\IEEEauthorrefmark{1}The University of Hong Kong, HKSAR, China
	\\\{mileswen, vli\}@eee.hku.hk}
\IEEEauthorblockA{\IEEEauthorrefmark{2}University of California, Berkeley, CA, USA
	\\\{arghandeh, vonmeier, poolla\}@berkeley.edu}
\thanks{This research is sponsored in part by the U.S. Deparment of Energy ARPA-E program (DE-AR0000340).}
}

\maketitle

\begin{abstract}
This paper proposes a novel phase identification method for distribution networks where phases can be severely unbalanced and insufficiently labeled. The analysis approach draws on data from high-precision phasor measurement units (micro-synchrophasors or uPMUs) for distribution systems. A key fact is that time-series voltage phasors taken from a distribution network show specific patterns regarding connected phases at measurement points. The algorithm is based on analyzing cross-correlations over voltage magnitudes along with phase angle differences on two candidate phases to be matched. If two measurement points are on the same phase, large positive voltage magnitude correlations and small voltage angle differences should be observed. The algorithm is initially validated using the IEEE 13-bus model, and subsequently with actual uPMU measurements on a 12-kV feeder.
\end{abstract}

\section{Introduction}
\label{sec:Intro}
Electric utilities often have limited or unreliable information to identify the phase (A, B, C) of connected loads. Moreover, phase moves due to restoration, reconfiguration and maintenance activities can happen frequently in distribution networks, and such changes are not always tracked continuously. Yet correct phase labeling is crucial in order to avoid excessive losses or reduced life of network components as a result of imbalanced loads, distributed energy resource integration, or insufficient harmonic mitigation. Wrong phase labeling is also a major source of error in diagnostic processes such as topology detection, state estimation~\cite{schenato2014bayesian}, and fault location ~\cite{vonMeier2014Chap}. To control and operate three-phase systems with active components, three-phase models are needed to adequately represent the network; however, validation of such models is not trivial because phase labels are often unknown~\cite{Stewart2014}. For lack of better information, utilities often assume their network is balanced and assign one-third of loads to each phase~\cite{Seal2011}. They may also use manual approaches for phase identification based on signal injection techniques. However, these solutions are not widely adopted in utilities, because they are costly, labor intensive, and error-prone. 

Fortunately, more and better measurement data are becoming available that can support reliable phase identification. While there was no need historically to monitor distribution networks very closely, the growth of diverse distributed energy resources motivates more intelligent monitoring strategies to observe and control the behavior of system components at greater resolution in space and time. The authors are collaborating on the development of one such advanced monitoring technology, micro-synchrophasors or uPMUs for distribution systems~\cite{vonMeier2014}, capable of providing very high-resolution, time-synchronized measurements of voltage magnitude and phase angles. Our proposed approach to phase identification employs this type of data, analyzing time-series measurements taken at different locations on a feeder. We are thus able to treat phase identification appropriately as a dynamic and time-variant problem. 

The proposed algorithm is shaped around the comparison of voltage magnitude and angle measurements at a pair of nodes on the network. The first step is based on the voltage magnitude correlation between different phases at the two nodes, using highly granular time-series measurements (120 samples per second).  However, these magnitude cross-correlations can still be masked by load variations and imbalances. To minimize the phase identification error, the second step of the algorithm compares voltage phase angle differences from uPMU measurements. Analytically, the combination of voltage magnitude and phase angle is entirely novel compared to other phase identification approaches. The proposed algorithm owes its strength to the precise time synchronization and sampling rate of the uPMUs, and the unprecedented ability to measure small phase angle differences on distribution feeders.

The rest of this paper is organized as follows: Section II describes related works. Section III explains the proposed algorithm. Section IV presents simulations and numerical studies, and Section V shows algorithm validation via actual measurement data from a uPMU field installation. Section VI suggests conclusions and future work.

\section{Related Works and Contributions }
\label{sec:Literatures}
Distribution networks are mostly unobservable beyond the substation. Switches and protective devices may not reliably communicate their status to the distribution operator, so the phase labeling and topology can only be determined with certainty by sending crews into the field. The knowledge of correct phase labels are crucial for different applications related to network topology, connectivity, and efficiency. In addition to the lack of observability, network unbalance, load variation, distributes energy resources, and frequent phase moves make the phase identification a challenging problem to solve. 

The available literature on phase identification is limited. In~\cite{Chen2011} a phase identification device is proposed to measure phase matches with point to point installations of the device, and ~\cite{Zhiyu2013} presents a signal injector device along with signal processing tool that can be used for phase identification. The disadvantages of these methods are the need for new hardware, communication links and trained staff to use them. Another class of approaches is based on load metering data. For example,~\cite{Dilek2002} uses power flow analysis and load statistical data to match phases in substation with aggregated loads. However, it implies statistical load information without considering load uncertainty.~\cite{Arya2011} has a similar objective as~\cite{Dilek2002}, presenting an optimization problem with load noise included. The approach is well-structured. Load metering data have to be time synchronized with a measurement device on each transformer, but this assumption still falls short of an actual load metering system. Moreover, the convergence of the proposed optimization is too sensitive to data quality. Voltage measurement cross-correlation for house meters data is presented in ~\cite{Pezeshki2012}. The method is easy to implement, but requires a three-phase load on each lateral as the reference, which is often unavailable.

A simple method based on voltage linear regression between meters and substation is proposed in ~\cite{Short2013}, using an aggregation of metering data to correlate with substation voltage. This method is sensitive to Geographical Information System (GIS) model inaccuracies. For phase identification purposes, the advanced metering infrastructure (AMI) voltage correlation in balanced feeders, feeders with PV resources, and feeders with inaccurate models can be error prone. Moreover, voltage measurements from AMI meters are typically available only in terms of hourly or 15 minute average values, where the averaging introduces additional errors in voltage regression-based methods for phase identification.

By contrast, the proposed method in this paper is based on a small number of micro-synchrophasors that provide high-precision voltage magnitude and phase angle measurements at up to 120 samples per second, taken at the secondary voltage level. Magnitude and angle measurements from a pair of locations are analyzed in two distinct steps to match each phase to its correct counterpart at the other location with a high degree of confidence. Phase identification could be one of many diagnostic applications supported by the relatively inexpensive uPMU instrument $(see: http://pqubepmu.com)$, and thus would not have to be its sole economic justification. 

The main contributions of this paper are as follow:
\begin{enumerate}
\item It presents a new technique for phase identification based on a combination of voltage magnitude cross-correlation and phase angle comparisons;

\item It validates the proposed algorithm on a 13-bus IEEE test feeder with synthetic phasor measurements and noisy load data, demonstrating that this optimization problem can be solved in a quick, simple and robust manner that yields unambiguous results; and

\item It further validates the algorithm with empirical measurements taken in the field on a 12-kV distribution feeder with prototype uPMU devices.
\end{enumerate}
\section{Phase Identification Algorithm and Analytical Tools}
\label{sec:IdenAlgl}

The phase identification process can be performed either on a one-time or an ongoing basis to assess the phase connectivity of customers or confirm the phase labeling for substations, laterals, or other network components. Time-series voltage measurements are available from uPMUs via a chosen medium (in our case, wireless communications) and are stored in a database for either continuous or on-request phase identifications. In general, three-phase voltage magnitudes along the feeder will differ based on load unbalance, independent voltage regulators operating on each phase, and single-phase distributed resources adoption. 

\begin{figure}[t]
\centering
\includegraphics[width=8.5cm,scale=0.7]{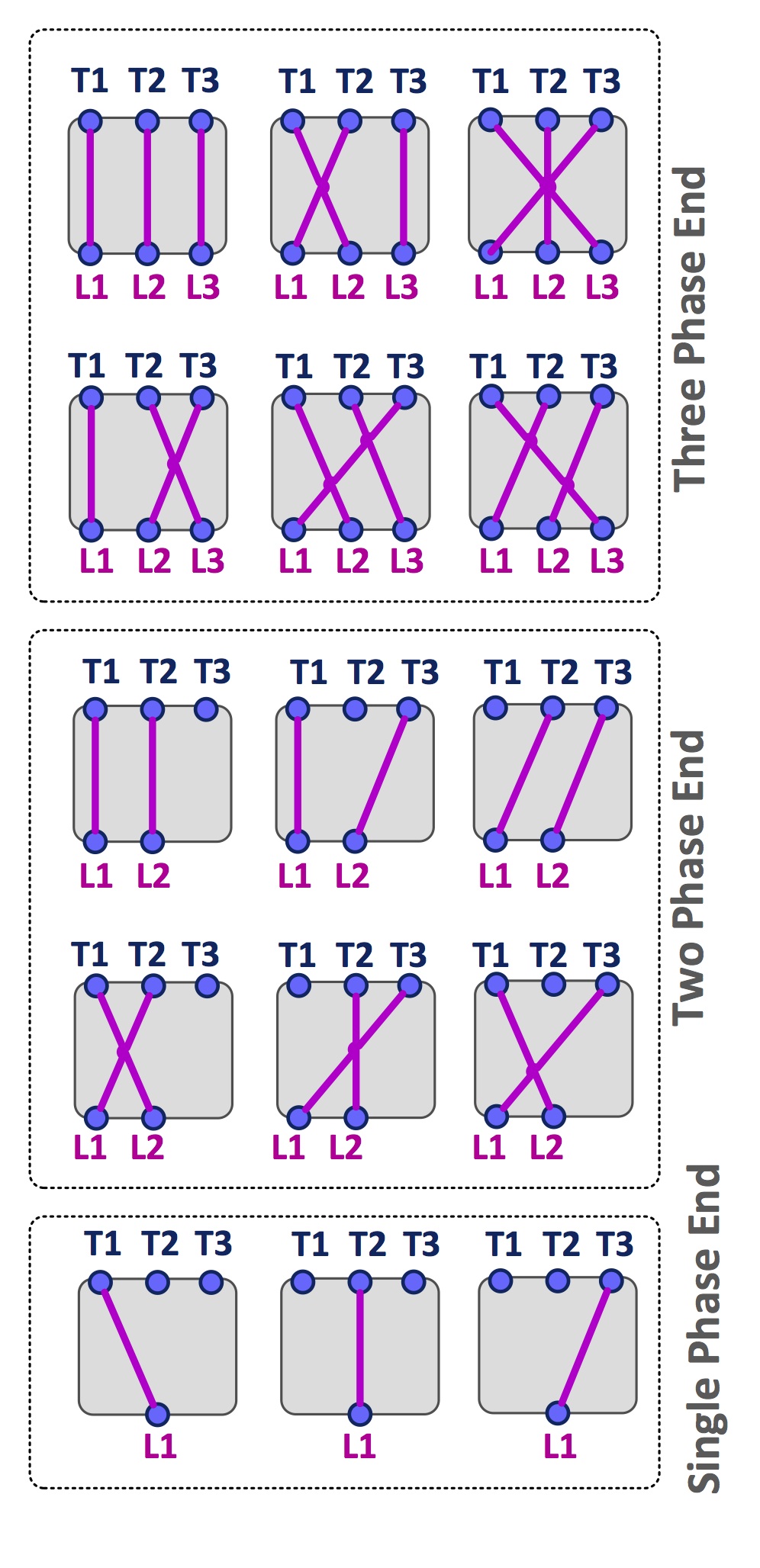}
\caption{Different possible configurations for a two bus connection.}
\label{fig:PhaseAllocation}
\end{figure}

Voltage correlation between two nodes on the same phase is an indication for phase matching. The hypothesis is that voltage at downstream nodes is correlated with the upstream node on the same phase. In the proposed algorithm, voltage correlation between two measurement locations are calculated for an initial phase allocation. The voltage magnitude correlation is sensitive to system balance, measurement uncertainty and load variation. Therefore, the proposed algorithm simultaneously tests the phase angle difference between different phases. In the case of multi-phase connections, 15 different combinations are possible for each pair. Figure~\ref{fig:PhaseAllocation} illustrates the different possible phase allocations for two ends of a three-phase line, where not all three phases may be present at the load.

The proposed algorithm is a brute-force search method based on the linear programming (LP) optimization structure. The objective function is to find the set of phase labels that yield maximum match with the measurements. Multi-phase voltage magnitude correlations and phase angle sequential differences have to be met simultaneously as constraints in the optimization problem. Following is the mathematical formulation for the phase identification algorithm:

Given a time series of $n$ voltage phasors, let $(V_1[t], \delta_1[t])$, $(V_2[t], \delta_2[t])$, and $(V_3[t], \delta_3[t])$ denote the time series voltage magnitudes and phase angles measurements on phase A, B, and C of the reference bus, respectively. Let $i,j$, and $k$, $i\neq j\neq k$ and $i,j,k\in\{1,2,3\}$, denote the phase labels at the target bus, respectively. Then the phase identification problem can be written as the following optimization problem:

\begin{align}
\argmax\limits_{i,j,k}\quad & \frac{\alpha F\big(V_i, V_j, V_k\big)}{n} + \displaystyle\sum\limits_{t=0}^n \frac{\beta G\big( \delta_i[t], \delta_j[t], \delta_k[t]\big)}{n}\nonumber\\
\mbox{s.t.} &\left\{\begin{array}{l}
F(V_i,V_j,V_k) = \displaystyle\sum\limits_{t=0}^n\Big( V_1[t]V_i[t]\\
\qquad +V_2[t]V_j[t]+ V_3[t]V_k[t]\Big)\\
G\big(\delta_i[t], \delta_j[t], \delta_k[t]\big) = \Big(\big|\delta_1[t]-\delta_i[t]\big|\\
\qquad+\big|\delta_2[t]-\delta_j[t]\big|+\big|\delta_3[t]-\delta_k[t]\big|\Big)\\
i,j,k\in\{1,2,3\}\\
i\neq j \neq k
\label{eqn:identification}\end{array}\right.,
\end{align} 
where $F(V_i, V_j, V_k)$ is the average voltage magnitude correlation function, $G\big(\delta_i[t], \delta_j[t], \delta_k[t]\big)$ is the average voltage angle shift function, and $\alpha>0$ and $\beta>0$ are the user-defined weights on the two objectives. Generally speaking, on the one hand, a larger value of $\alpha$ should be used if the system is expected to be heavily unbalanced and, on the other hand, a larger value of $\beta$ is needed if the two buses being investigated are adjacent to each other.

The proposed algorithm is fed and validated with uPMU voltage phasor measurements. However, note that it can be implemented on current measurements or even a combination of voltage and current measurements. In practice, obtaining voltage measurements is often easier due to the installation logistics of potential and current transformers. Also, the uPMU can obtain single-phase voltage phasor measurements simply by plugging into a standard 120V outlet. 

Because the algorithm is pairwise, it is scalable to any number of network nodes or installed uPMU devices.
\section{Simulations and Numerical Study}
\label{para:sim_results}

\begin{figure}[t]
\centering
\includegraphics[width=8.5cm]{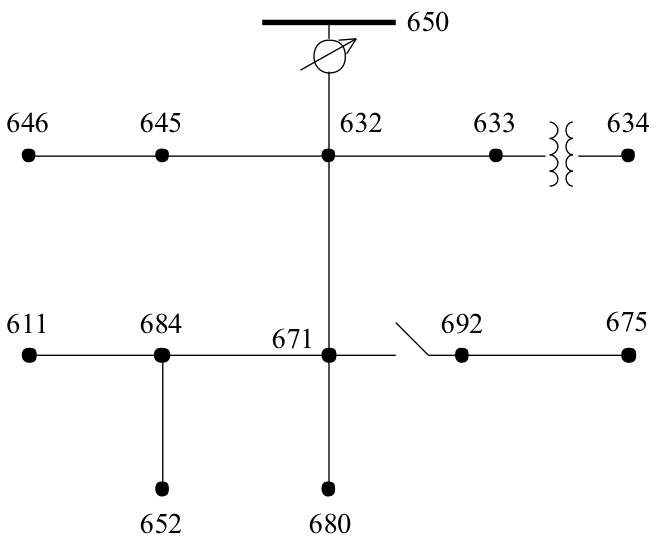}
\caption{IEEE PES 13-bus feeder test system~\cite{IEEE13node}.}
\label{fig:13node}
\end{figure}

\begin{figure*}[htbp]\centering
	\begin{subfigure}[t]{8.9cm}\centering
		\includegraphics[width=8.9cm]{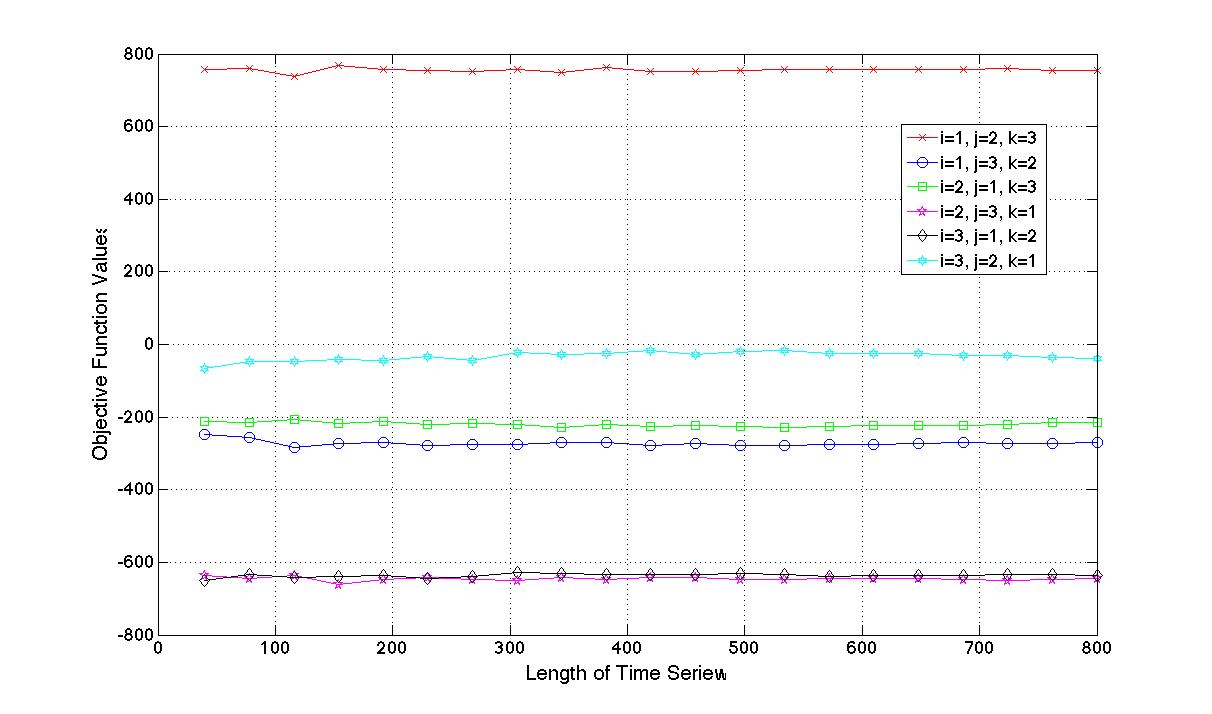}
		\caption{Phase identification results between Bus 632 and Bus 633.}\label{fig:sim632633}
	\end{subfigure} 
	\begin{subfigure}[t]{8.9cm}\centering
		\includegraphics[width=8.9cm]{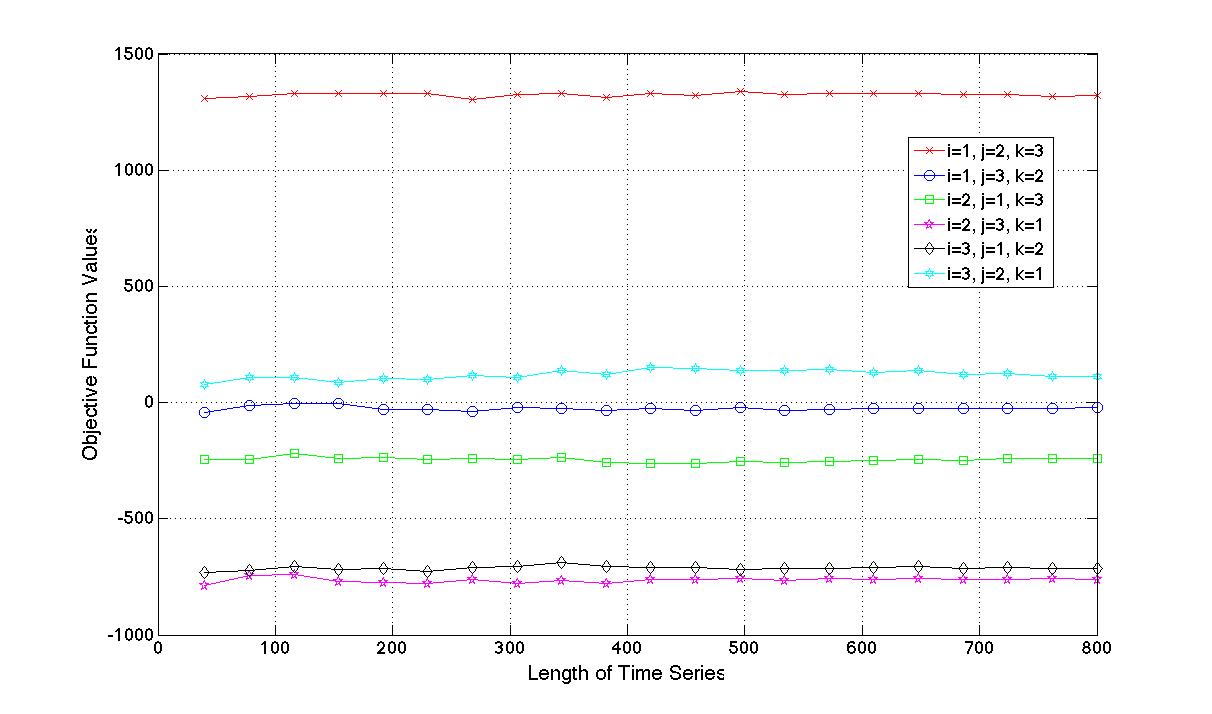}
		\caption{Phase identification results between Bus 632 and Bus 671.}\label{fig:sim632671}
	\end{subfigure}
	\begin{subfigure}[t]{8.9cm}\centering
		\includegraphics[width=8.9cm]{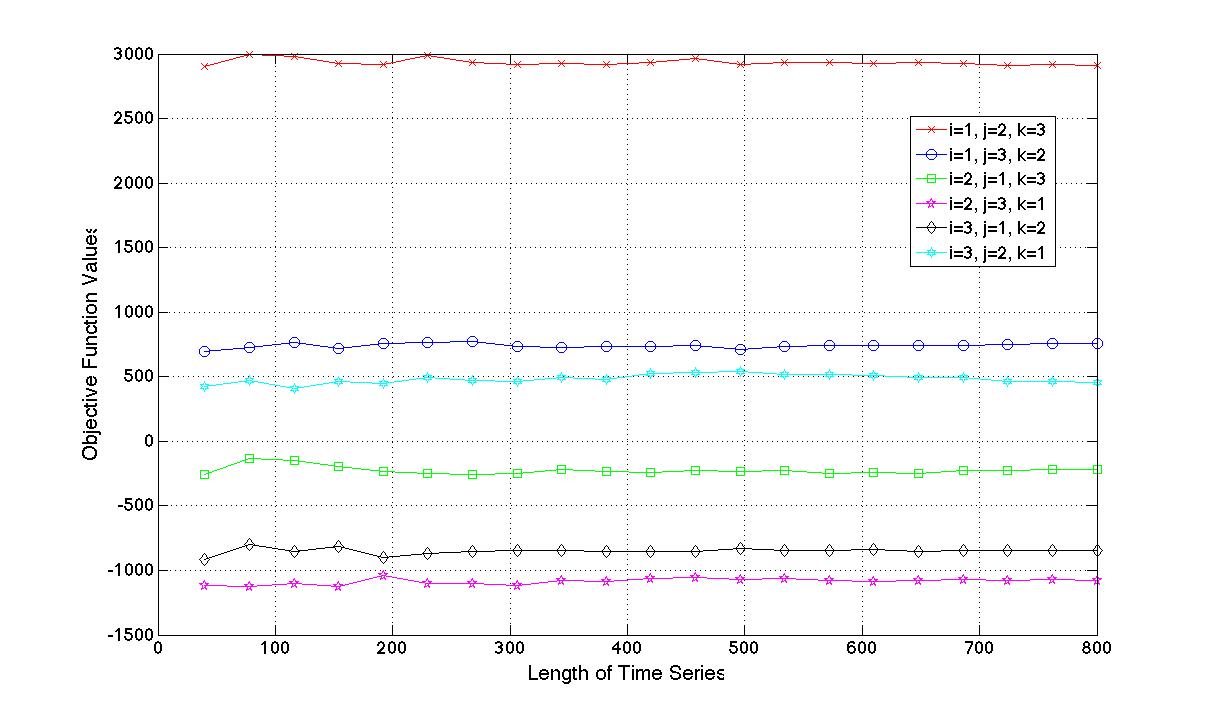}
		\caption{Phase identification results between Bus 671 and Bus 675.}\label{fig:sim671675}
	\end{subfigure}
	\begin{subfigure}[t]{8.9cm}\centering
		\includegraphics[width=8.9cm]{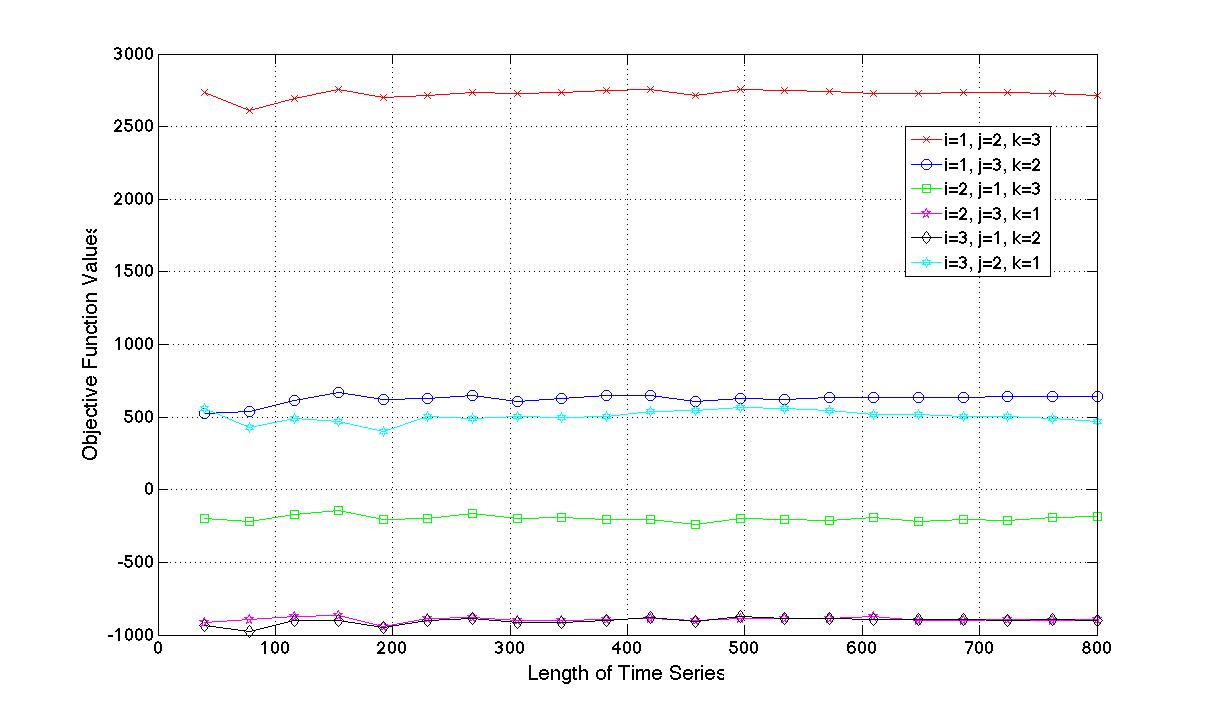}
		\caption{Phase identification results between Bus 671 and Bus 680.}\label{fig:sim671680}
	\end{subfigure}
	\caption{Phase identification results between selected nodes in the IEEE PES 13-bus feeder test system.}\label{fig:resultsSim}
\end{figure*}
To test our proposed phase identification method, we carried out a numerical study by simulating the IEEE 13-bus feeder test system, as shown in  Fig.~\ref{fig:13node}, using the OpenDSS simulation environment ~\cite{openDSS}. We first generate random load profiles for each bus in the system and then continuously run one thousand power flow computations on the system, from which we obtain a series of voltage phasors on all buses. These synthetic data are used to represent the time-series of voltage phasors collected by uPMUs. We then feed the synthetic uPMU data into MATLAB, where our phase identification algorithm lives.

For this particular distribution network, we set $\alpha=1$ and $\beta=1$. Our further investigations reveal that, since our simulated data are so clean, the values of $\alpha$ and $\beta$ will not change the final conclusions.

The phase identification results between Bus 632 and Bus 633, Bus 632 and Bus 671, Bus 671 and Bus 675, and Bus 671 and Bus 680 are shown in Fig.~\ref{fig:resultsSim}. In each case, we iteratively compute the results 21 times using different lengths of time- series data. As we can see, in each phase labeling scenario, the case when i = 1, j = 2, and k = 3 attains significantly larger objective function values than others.

The results indicate that, for each case, the objective functions are consistently maximized by having $i=1, j=2$, and $k=3$. In other words, all the buses that we have studied have their phases labeled correctly, which is indeed true for this test system.
\section{Field Test Results with micro-PMU Data}
\label{para:actual_data}

\begin{figure}[t]
\centering\includegraphics[width=8.5cm]{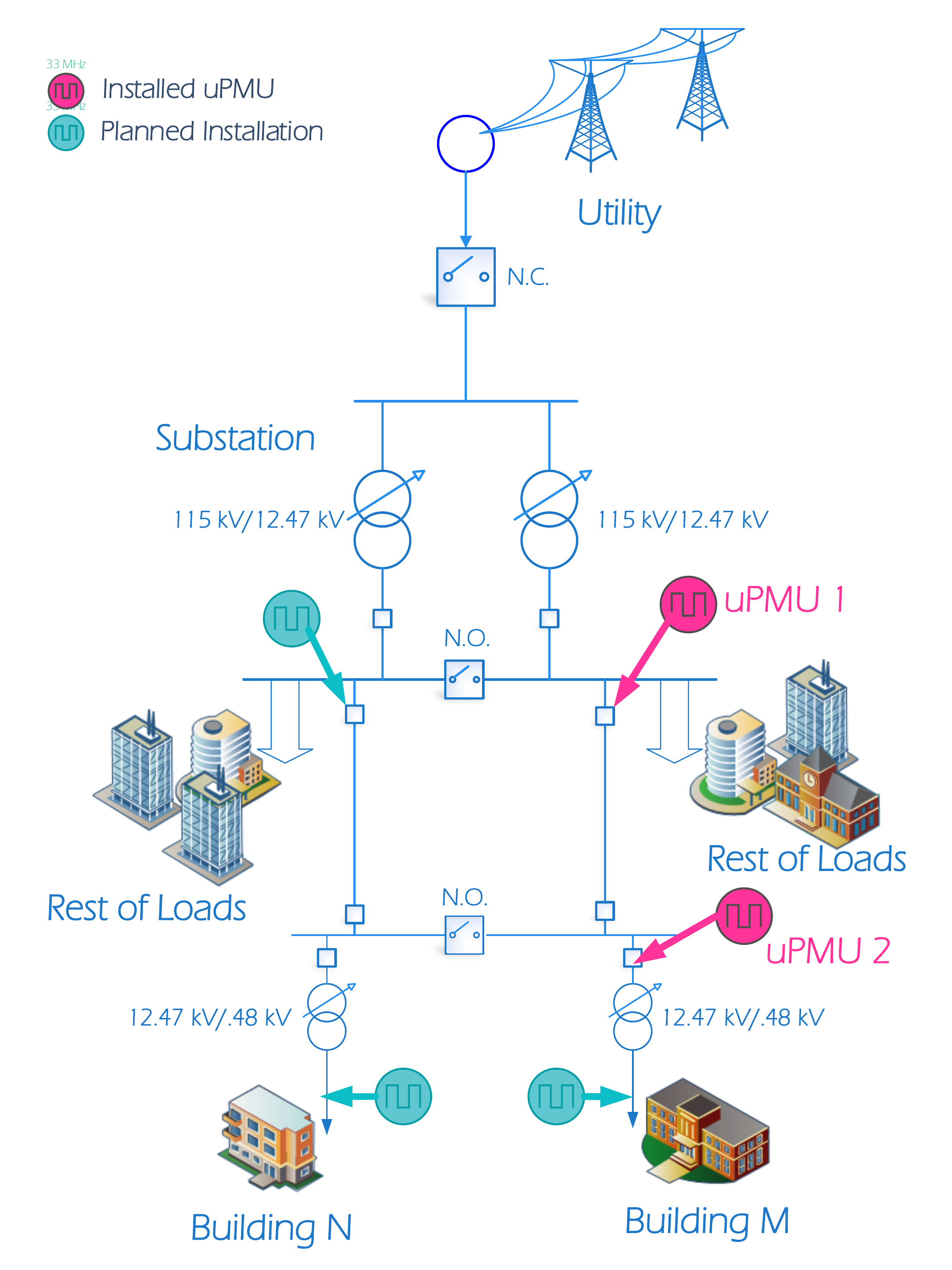}
\caption{Simplified Diagram of the Case Study with installed uPMU and planned uPMU for future installations.}
\label{fig:CaseStudyMap}
\end{figure}

As part of an ARPA-E funded effort, the authors are collaborating with Power Standards Lab and Lawrence Berkeley National Lab (LBNL) to install a number of prototype uPMU measurement devices at various distribution system locations. We choose data from the pilot installation at the UC LBNL campus, which includes several buildings with research and educational functions, to validate the proposed algorithm. Fig.~\ref{fig:CaseStudyMap} shows the abstract line diagram for the case study. Phase identification is done between the substation and Building M. The data set applied for algorithm validation includes 120 samples per second for a period of one-hour, collected by two uPMUs from 08:00:00am to 08:59:59 on 10 July 2014. 
Since real-world data are very noisy compared with the simulated data, the voltage magnitude correlation values, namely, the value of $F(V_i, V_j, V_k)$, are significantly smaller than voltage angle difference values, namely, the value of $G\big( \delta_i[t], \delta_j[t], \delta_k[t]\big)$. Therefore, in order not to have the objective function values in (\ref{eqn:identification}) dominated by a single term, we give a higher value to $\alpha$ by setting it equal to $10,000$ while keeping $\beta=1$. 

\begin{figure}[htbp]\centering
	\includegraphics[width=8.5cm]{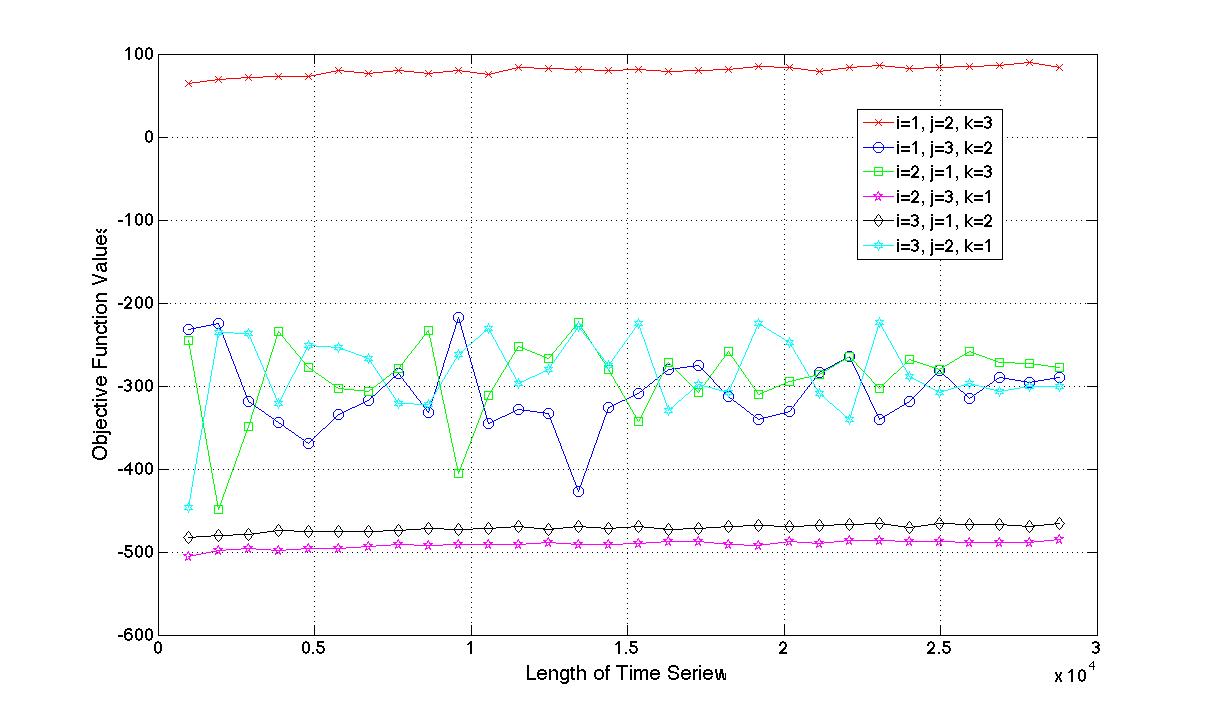}
\caption{Comparing volage phasors on two adjacent buses.}\label{fig:fieldtest}
\end{figure}

As shown in Fig.~\ref{fig:fieldtest}, the scenario when $i=1,j=2,k=3$, namely, matching phase A, B, and C of one bus with phase A, B, and C gives us the highest objective function values. From this result, we are confident to say that the two uPMUs that we have installed are correctly labeled.

\section{Conclusions and Future Work}
\label{sec:conclusions}

In this paper, we proposed a novel method for phase identification in distribution networks based on micro-synchrophasor (uPMU) measurements. Our method works based on the observations that, in an unbalanced three-phase system, time-series voltage magnitudes on two ends of the same phase should show significantly stronger correlations that those on two ends of different phases. Furthermore, after excluding Delta-Wye transformer phase shifts, the voltage angle differences should be very small for measurements on two ends of the same phase. We tested our method both by simulating the IEEE 13-node test feeder system and using field measurements from two uPMUs. 

As mentioned in Section~\ref{sec:IdenAlgl}, phase identification methods solely based on voltage magnitude correlations will fail to converge when the system is well balanced. On the other hand, methods based on voltage angle differences cannot be trusted when more than one delta-wye transformer is present on the line, since phase shifts in multiples of 30 will no longer be uniquely traceable. These scenarios will be investigated in our future work. We look forward to continuing this work with the installation of additional uPMUs at the campus pilot site and other locations in collaboration with several utilities. This will also allow us to examine the impact of uPMU locations along with the characteristics of different distribution networks.

\section{Acknowledgment}
\label{sec:Acknowledgment}

We thank our colleagues on the project team including: Merwin Brown, Lloyd Cibulka, Dan Arnold, Laura Mehrmanesh, David Culler, Michael Andersen, Alex McEachern, Thomas Pua , Emma Stewart, Sila Kiliccote, and Charles McParland.

\bibliographystyle{IEEEtran}
\bibliography{bibTex_PhaseIden}

\end{document}